On locating statistics in the world of finding out

*Chris J. Wild*

*Department of Statistics, The University of Auckland*

*July 20, 2015*

**1. Introduction**

"An unexamined life is not worth living," said Socrates as his own life drew to a tragic close. Even more emphatically, an unexamined education is not worth delivering. Periodically, we need to examine what we are doing and why — starting from the most fundamental of fundamentals. Ograjenšek and Gal (2015, hereafter O&G) state, "purpose, motivation, and skill transfer ultimately determine the process of acquiring, understanding, and connecting the knowledge learned within the classroom to the roles that statistics may play in learners' lives beyond the classroom walls."

They are right to urge that we focus on purpose, motivation and skill transfer, which I prefer to think of as capability. *Purpose* ("What's it for?" "How does this fit into the larger scheme of things?") is a necessary ingredient of understanding, helps with motivation, and is a prerequisite for gaining real-world capabilities. *Motivation* helps students answer in their own minds "How does this advance what I care about?", or more crassly, "What's in it for me?" *Capabilities* steers us to think about how our education processes can be tuned to maximise the chances of potential benefits being realised at future points of need in students' lives. This really is getting back to basics; or at least it would have been if many of us had actually ever been there before!

For statistics majors and graduate students, the core motivation probably comes down to interesting programmes involving doing things that they're relatively good at, and leading on to interesting, important and relatively well-rewarded careers. Here, we roughly know what's needed even if the implementations could be improved. What I want to think about is motivating those who take some statistics (e.g. an intro course) *to want to learn more* at some stage in their lives. What, potentially, is in it for them? How can we ground statistics within central life-concerns?

Perhaps the grounding of statistics in "real life" could start with how we learn and how we project from our experiences – with the learning processes that form the internal world-views that inform all of our decisions and actions throughout life. We all bemoan the fact that large numbers of students show little interest in anything that is not going to be assessed, even to the extent of making the minimum efforts "to get through", so brilliantly encapsulated in the snide, "C's get degrees" (C is the minimum passing grade in our system). It is as though all that matters are the short-term horizons of paper qualifications (diploma and degree certificates). How can we turn this around? Conversations about some realities of lifetime career success could help. They might proceed along the lines of the three paragraphs that follow.

"Paper qualifications help you get that first-job foot-in-the-door, but as soon as you are inside they become irrelevant. Scaling the stairway to success depends on what you know, who you know and what you can do – your capabilities. Critical among them are how well you can learn, identify and solve problems, make decisions, communicate (including negotiate and persuade), and network. Engaging seriously with statistics can help you improve all but the last.

The most important ingredient for making good decisions is a sound knowledge or intuition about the way things work in the world you are operating in. Most of the specific facts you cram into your brain for exams will quickly be forgotten or go out of date. What can stick in your brain and work for you are broad principles and an ability to operate from them. It is important then to be able to quickly acquire specific knowledge when you need it, to integrate it with the other things you know, and then to apply it. There are ways in which statistics enables you to think about the world and the information you are getting that you would be incapable of otherwise. It also helps you obtain, critically evaluate, and communicate *evidence* about the ways things work in whatever world you work in.

Unfortunately, however, the brain is like a set of muscles. There is no easy way to becoming a world-class athlete. You need to do intensive physical training. Similarly, the more you exercise your brain the stronger it gets and the more things it becomes capable of doing (by building new circuitry). If you keep taking the easy way out your brain remains limp, weak and slow. What is most critical is not what facts you can remember from your studies, it is about how capable your brain is of processing new information and coming up with the questions and answers that matter."

Most of our beliefs about how the world around us operates come from opportunistic, "anecdotal", impression-gathering processes ("System 1" in Kahneman, 2011) fed by what we see, hear and read in our every-day lives. This "informs" the vast majority of the action-decisions we make; fast, largely effortless decision-making processes based on "gut feelings". These processes have served us very well evolutionally. But often what we think we know is not actually correct. When we deem our knowledge bases may be inadequate, or where consequences of getting it wrong are sufficiently high (and), it is worth making a conscious effort to find out more.

In the galaxy of forming views about how the world around us operates, statistical inquiry is a tectonic plate on the planet "Purposefully Finding Out". Neighbouring plates include "sourcing from 'trusted others'" (people, books, the internet) and "Qualitative Research methods" (Petocz and Reid, 2010; Gal and Irena Ograjenšek, 2010). How do they interrelate? Statistical enquiry ("quantitative research methods") and "qualitative research methods" are expensive, organised, principled forms of knowledge gathering, though the latter appears to have been developed primarily for research about people and societies. They have complementary roles in satisfying a need to know. But like tectonic plates they have areas of overlap and sometimes destructive energy is unleashed when they rub up against one another. We conjecture this is mainly the result of inadequate communication and misunderstanding. Ironically, much if not most research in statistics education research (and education more generally) is actually qualitative; see Groth (2010).

**2. Scouts and Surveyors**

I asked Andrew Balemi, a colleague who has worked in marketing research for many years, how these things work in the real world of market research. His take on it is this.

"*I think of the qualitative researchers as the scouts and the quants as the surveyors.* They (the quals) go out exploring the frontier and come back and say, 'There's a hill over there, there's a ravine and a plain over there …' The quanties come through and say, 'The hill is 200 metres high, the valley is not as extensive as we thought, and …'"

You cannot actually do quantitative research (statistical investigation) until you know quite a lot about the area under study, for example have some clarity (rather than vague feelings) about issues that need to be addressed and a very good idea about what to record and how. If you don't then you need to do some sort of precursor research. But how do you identify the issues that really matter? How do you crystallise precise questions from vague intimations and desires to know? How do you determine what to record and how? At least for research on people, qualitative research-methods provide systemised methods for making progress on these aspects and more. In that way it can serve as a precursor to quantitative research. Its methods critically leverage the ability of human beings to notice things, to "read" other people, and even simple serendipity. Close observation may lead to researchers noticing things they never would have thought to look for a priori. (Here we are talking about "noticing things in the world" in contrast to the statistical "noticing things in data.") Whereas in statistics we *use* types of measure and categorisations that have already been determined, much of qualitative research appears aimed at *discovering* or developing fruitful and informative categorisations.

Before going further I want to place in the background of this discussion Jock MacKay and Wayne Oldford's PPDAC model of the statistical inquiry cycle from the early 1990s. PPDAC stands for **P**roblem-**P**lan-**D**ata-**A**nalysis-**C**onclusions. (This is unpacked by Wild and Pfannkuch, 1999 and MacKay and Oldford, 2000 with minor differences in emphasis.) What appeals to me about the PPDAC model, and puts it ahead of competing descriptions of statistical inquiry, is its emphasis on the up-front decision-making elements of inquiry, particularly by breaking out the initial "P" and thus spotlighting problem (or question) crystallisation. The second P (Plan) is up-front decision-making elements, which includes choosing measures, categorisations and observational schemes. Questions, measures and other up-front elements of PPDAC do not spring forth into the world fully formed like the goddess Athena from the head of her father Zeus (though it would be hard to detect that from most statistics instruction). They are the result of prior awarenesses, thought and investigation much of it qualitative. Students do need to have experiences in wrestling with the upfront elements of PPDAC and know there are tools available which can help. And "*Garbage in, Garbage out*" should be burned indelibly into the brain circuitry of every statistics students. In my recent MOOC, I characterised this as "the first law of data analysis". It applies right from the beginning of statistical investigation where bad initial decisions can doom the entire enterprise.

From O&G's description of qualitative research methods, their main value to statistical inquiry would seem to be in improving the development of Problem and Plan. O&G also talk about contributions to Conclusions. In statistical terms, the form of those O&G show is idea or hypothesis generation about why the results turned out the way they did and their implications. This is clearly crucial to better understanding our world, decision making, and planning future inquiry.

We will now compare and contrast qualitative and quantitative approaches under a number of headings.

*Context*

In their Section 3, O&G state that in qualitative research things "cannot be fully understood without considering the context in which people live and function." It almost suggests that this is a defining distinction between qualitative and quantitative approaches. But quantitative researchers are often interested in when relationships between variables of interest change depending on values of variables that capture elements of the background context (we look for interactions with context-capturing variables)

The essential difference, I think, is that in statistical inquiry we assume that the variables we have decided to record capture the essential elements of the context whereas qualitative research is open to these elements becoming evident during the process of observation. This dichotomy is a little too simplistic. Qualitative "data" such as hours of video footage or interview transcripts are rich enough to permit elements one would never have thought to purposefully record to come to notice during analysis. So qualitative research tends to "over-capture" data on a smaller number of things (err on the side of taking much more than you think you will need) whereas quantitative research tends to (probably under-) capture a much smaller amount of data on a large number of things.

*Looking out for the unexpected*

Both communities are constantly on the lookout for the unexpected but with more emphasis on different stages of the process. We statisticians tend primarily to look out for the unexpected in the relationship-patterns between our pre-defined variables and discrepancies between data and statistical models. Qualitative researchers are on the lookout for things that jump out as "unexpected" or interesting against the background of their life's experience in observational processes performed before the "variables to record" have even been defined. They are also interested in interactions (in the statistical sense).

*"The Principle of Suspicion"*

Both communities worry about mistaking artefacts for facts. They both worry, "Is what I think I see really there?" Consequently both work on fostering critical instincts and developing protective techniques. Klein and Meyer (1999) list 7 principles for interpretive studies. The last they call "the principle of suspicion". It's a brilliant phrase. We statisticians should steal it! What differs between the communities is the spectrum of artefacts its members are most sensitised to protecting against. We will elaborate in the next paragraph.

*Inference and representativeness*

O&G talk about "inference" and state that "both qualitative and quantitative researchers make inferences." We may be hearing some faint echoes from the Tower of Babel here. The qualitative and quantitative communities seem to be using the same word for two critically-important-but-very-different types of endeavour. What qualitative techniques like triangulation address are inference in the sense of "Is what I think I am seeing really there?" through "Are we all seeing the same things here?" ("Here" being the setting/people we are currently observing). In statistical terms these are really "measurement" issues; issues of how can we record what is going on or validly categorise what we see.

But this is not "inference" in a statistical sense. Statistical inference comes after this. It operates from an assumption that useful categorisations and measurements are being used and then goes on (in the categorical case) to address the commonness or relative-commonness of categories extrapolating beyond who or what has been observed to some wider population or process. The stories we have been given are about coming to valid (or at least useful) ways of recording what's happening for the subjects under study and intimations about generalisations to what might be happening more generally (scouting). Classical "statistical inference" is about generalising from "sample" to population or process reasonably reliably (surveying). It is much more confirmatory.

The qualitative community recognises and confronts the sometimes ephemeral, subjective/personal, value-encoding and contentious nature of much "classification and measurement". But because of its intensive and detailed nature, the per-subject cost of qualitative research tends to be very high leading to the use of small numbers of subjects. One thing that all statisticians know is that, apart from things that virtually always or never happen, you cannot reliably tell anything useful about relative common-ness of categories from small numbers of subjects. Thus while detailed investigation of small numbers may often be excellent for uncovering issues, it cannot address representativeness. Transferability is another issue. Whether you can transfer findings from a population or process studied to one that was not is not something that is justified by statistical research design. It needs other forms of argument. O&G say that "qualitative methods go beyond exploratory processes" but their further discussion only really talks of "to conjecture about underlying influential factors thus helping the creation of the theoretical framework for hypothesis testing." In statistical language this is still exploratory. It is still setting up further investigation. It is not confirmatory.

We might capture the contrast with the following, admittedly oversimplified, dichotomy. Qualitative research focusses more on getting the recording of observation right, whereas quantitative resesearch focusses more on getting extrapolation right. Qualitative researchers seem to be much more sensitised to protecting against artefacts in measurement and categorisation and have tools for protecting against this. Statisticians are more sensitised to look out for biases in selection processes and over-claiming when generalising from data on subsets to the behaviour of populations or processes more broadly.

All of these things we are hearing about qualitative research seem to me to be (i) very important and (ii) largely complementary to the things statisticians generally worry about and the ways they worry about them. Both the qualitative and quantitative research communities are made up of smart, sensible people doing smart, sensible things that address critical research imperatives but who have yet to really figure out how to talk to one another without talking at cross purposes. There is room for a lot of work on gaining cross-community clarity about the nature of the research issues being addressed and how they are addressed. Attempts to carefully translate from one community's use of language to the others may be a useful way forward.

### 3. What should we do in intro stat?

By intro stat we mean the common university-level introductory statistics course for the masses – still probably the discipline's most visible (if ill-utilised) shop window. Beyond intro stat, I see definite merits to encouraging qualitative electives for majors heading for employment or research destinations which will involve finding out about the thinking, motivations, emotional responses, decision making, behaviour and social organisation of people. But here, I am going to concentrate on intro stat.

*Realities and Goals*

"The curriculum of the intro course is already overcrowded." It is a commonly-heard refrain whenever we are asked to include something new. There are lots of different domains from which we feel we should be including more (e.g. aspects of data science). We obviously can't keep adding without subtracting. What should we pull out? …

A forthcoming article in the *International Statistical Review* by Jim Ridgway is entitled, "*Implications of the data revolution for statistics education*". As I write this I am also drafting my discussion of a paper by George Cobb entitled, "*Mere Renovation is Too Little Too Late: We need to rethink our undergraduate curriculum from the ground up*" (to appear in *The American Statistician*). It is a song I have also been singing, particularly in regard to the intro course. With the advent of big data and the opportunities and challenges from "data science" the standard intro course has passed its use-by date. It reveals far too little of the exploding world of data and does it far too slowly. We really do have to rethink it from the ground up, radically rebalancing its priorities. So it is timely to also throw the up-front elements of investigation, including qualitative methods, into the mix being considered.

But even the often-stated goals of the conventional intro course are impossible to deliver in a one semester. Let us stop trying to fool ourselves. The stated goals are really just directions in which we want to improve capabilities, we are never going to be able to *perfect* those capabilities. We should abandon the self-deluding fiction that we can build much in the way of capabilities of fundamental and lasting value in a single course. We can only hope to build competence in a small number of relatively small things. If we are serious about fostering future capabilities, the best we can hope for is to initiate growth processes by planting crucial seeds and trying to ensure that they fall on fertile ground where they can grow up and bear fruit over time as and where they are needed. The truth is that to do anything nontrivial, and do it well, the students will have to learn more. Our job is to plant the seeds that will cause them to recognise where statistics might help them in their work (or wider lives) and, having realised that, to recognise their needs for further, just-in-time learning – together with some ideas about how to find these things. (The better alternative, for those who are in a position to be able to do so, may be to bring in someone else with the appropriate skills.)

So we should not think in terms of "the single course which may be the only instruction you'll ever have". It might be the only *formal* instruction you'll ever have but it should plant the seeds for a realisation of "additional learning I need" – to pop up at the time of need. This is not just an intro-stat thing. It is impossible to teach everything needed to solve practical problems in a limited number of courses. If we are focussing on "capabilities", all learning should try to plant seeds that will later trigger a realisation of needs for further learning about critical related-issues the course(s) has not had time to teach.

So if we are going to expose intro statistics students to more and faster, what has to give? Details! This is not an enormous loss because the details we teach are quickly forgotten and most go out of date quickly as well. Additionally, anything we do has to be done extraordinarily efficiently because intro-stat is so overburdened by competing imperatives.

*Motivation and capability, awareness and competence*

From the vantage point of motivation, I think it critical that intro stat prioritise, "What I can do with data and what data can do for me" to get compelling buy-in (I would go further and say this should be the highest priority; that doing this well would plant the most fertile seeds possible for a desire to learn more now and in the future.) This requires prioritising a breadth-of-awareness of things we can do with data. It also requires prioritising students gaining the confidence to jump into data themselves and beginning to explore. Breadth of awareness (of what is possible and what to worry about) needs to be much more extensive than the breadth of competence. Peripheral awarenesses can trigger "I think I can do something about this. I've seen something a bit like it before" and point to, "There are some things I need to find out about before I proceed." Because of the speed of fading memory, unless the time gaps from instruction are very short indeed, the details needed for competence will have to be obtained later, close to the point of need, anyway.

Faced with self-acknowledged knowledge or competence gaps, to channel the 80s movie Ghostbusters, "Who you gonna call?" One of my goals for statistics service teaching into other disciplines is to help students recognise, in their future lives, when they should call in a statistician. But they, and statisticians, should also recognise when it might be highly beneficial to call in someone with skills in qualitative research.

*Sense making*

The distinctions made in applied statistics teaching tend to be aimed at decisions about analysis tools and pathways and not about more fundamental sense making (Grolemund and Wickham, 2014). An analysis can only help us make sense of the real world if the questions being asked make sense to us, the variables it uses – the measures and categories – themselves make sense in terms of the questions, and what is uncovered in the data makes sense in terms of the questions and wider aspects of the context reality. It is only when statistics stops being arcane magical incantations and starts being incorporated into a person's stream of "commonsense" and sense making that it can reliably be summoned up and used well in real life. It is imperative that a need for these things to make sense become ingrained in statistics students. For most it currently hasn't. They're too busy playing what they think is the real game, pulling the "right" tool out of their tool bag to "get the right answer". (Hint: It's almost certainly the one you've just learned!) Perhaps it is because students spend so much time having to do things outside their competence zone under time pressure that many seem to become almost comfortable with operating without things making sense. It then becomes their norm and they lose the drive to doggedly persist until things do make sense. This is not just an intro-stat issue. It is pervasive.

*"Ask good questions"*

Leading statistics educator Allan Rossman gives a great talk entitled "Ask Good Questions". It is aimed at teachers of statistics improving their teaching but it supplies a slogan with a much wider currency. The key to (effectively) instigating creative or critical thought about almost anything is to *ask oneself* good questions (see the discussion of "trigger questions" and "worry questions" in Section 4 of Wild and Pfannkuch, 1999, as a means of stimulating useful lines of thought).

Downplaying the "give a man a hammer and every problem looks like a nail" truism, certain types of questions push well-informed investigators towards qualitative methods as a source of tools to help find answers. Given the paucity of time available in intro stat perhaps rather than describing what different qualitative methods do, perhaps we should simply seek to improve the propensity for our students to ask those sorts of up-front questions.

Good questions can help us to crystallise a need to know, and good questions can initiate thinking about how we could come to know that. This may be the best we can hope for from intro stat, and it is certainly not an objective to be sniffed at. What is my need to know? How could I find out about that? What is the question here, really? Do these measures truly address those questions? Are they biased in some way? … This may lend itself more to classroom brainstorming than formal instruction as a way of forming valuable habits of mind.

I like O&G's focus on driving everything off, "What is my need to know?" We can then turn attention to, "How could we find out?" PPDAC is a good starting structure for organising thinking. Case studies addressing questions like O&G's "Why do customers stay with a service provider after a service failure?" are needed as problems to brainstorm about and in this discussion qualitative techniques like focus groups can turn up naturally as aids to making progress on, "How we could we find out about that?" Giving attention to "Where do ideas come from?" is important. Ignoring this consigns statisticians to being mere bit players in someone else's drama.

Subjective/objective is not the right dichotomy for encapsulating differences between qualitative and quantitative research. Quantitative research is never entirely objective. There are subjective decisions made qualitatively all along the PPDAC path starting with problem isolation and measurement. Any of them can introduce unconscious biases. And good qualitative research worries about objectivity. Openness to the unexpected during data collection versus concerns for representativeness is a better one. But there is a continuum that has qualitative research at one extreme and extends through mainstream statistics into machine learning at the other extreme. It is a continuum based on the levels of human engagement (leveraging human capacities to notice and human direction of processes) versus automation. The boundaries are nowhere sharply defined. All good scouting includes elements of surveying. All good surveying includes elements of scouting. But we need a simple, understandable, starting dichotomy to get the discussion started, a dichotomy that can be nuanced later. I think Balemi's "scouts and surveyors" does this rather well.

Not only are the boundaries not sharply defined, they are blurring faster by the day as video footage and audio transcripts, previously largely the province of qualitative researchers, are being attacked using quantitative approaches. For example, in addition to the usual numeric and categorical variables, some predictive analytics software now allows free-text fields as input fields from which the software will try to automatically derive variables using text analytics that are then fed as into their suites of predictive models as additional predictor variables.

So let's come back to the question we headed this section with, "What should we do with intro stat?" I return to my vision that a modern intro stat needs to convey very quickly a broad appreciation for what you can do with data and statistical investigation to show that this is an area that can convey real benefits "for me". Continuing with small views of the subject dooms us to the current situation in which great swathes of the student cohort leave our discipline's most visible shop window with a view of statistics and statisticians as doing "hard boring stuff of little relevance to the real world I will inhabit" (or worse still, not even hard, just boring!).

Needing to convey broad appreciations quickly means that everything must be conveyed very efficiently. The biggest benefits from the fewest bucks with regard to the qualitative methods come, I believe, from teaching our students to ask good questions at key stages of an investigation, or about key stages of a reported investigation. The contexts that are easiest for students to think about are those that connect with their life's experiences and the most common of those involve people. Then, "how could I find out about that?" should often give opportunities to talk about qualitative techniques. But all of this is useless for capability building unless students can be habituated into coming up with these types of questions *for themselves*. So there has to be a medium for sufficient practice to establish habit. I think making group brainstorming a routine part of how we start thinking about things can supply such a medium. We also need a supply of case studies or scenarios where it is easy to do such thinking and sharpen both the scouting and the surveying instincts. And, in PPDAC with its additional prompts to up-front thinking, we already have a good scaffold on which to tie it all together.

I thank Ograjenšek and Gal for a very stimulating paper that provides a great deal of food for thought.